\documentclass[proc_ichep08_slac_one]{revtex4}   %%FRK: renamed slac_one.rtx; v3.04
\usepackage{graphicx}
\usepackage{fancyhdr}
\usepackage{latexsym,amssymb,amsmath,amsfonts,epsfig,graphics,xspace} %%FRK
\newcommand{\DS}[1]{$\mathsf{#1}$\xspace}       %%discrete symmetry

\newcommand{\bdi}{\begin{displaymath}}
\newcommand{\edi}{\end{displaymath}}
\newcommand{\bfi}{\begin{figure}}
\newcommand{\efi}{\end{figure}}

\newcommand{\beq}{\begin{equation}}
\newcommand{\eeq}{\end{equation}}
\newcommand{\beqa}{\begin{eqnarray}}
\newcommand{\eeqa}{\end{eqnarray}}

\newcommand{\id}{\mathrm{d}}                    % integral measure d
\newcommand{\ii}{\mathrm{i}}                    % imaginary i
\newcommand{\vecbf}[1]{\boldsymbol{\mathrm{#1}}}% bold vectors
\newcommand{\st}     {spacetime}

\newcommand{\half}   {{\textstyle \frac{1}{2}}}

\begin{document}

\title{New Indirect Bounds on Lorentz Violation in the Photon Sector}

\author{F.R. Klinkhamer}
%\email{frans.klinkhamer@physik.uni-karlsruhe.de} %%FRK
\affiliation{Institute for Theoretical Physics, University of Karlsruhe (TH),
76128 Karlsruhe, Germany}

\begin{abstract}
Direct laboratory bounds on the 9 nonbirefringent Lorentz-violating
dimensionless parameters of modified-Maxwell theory range from the
$10^{-7}$ level to the $10^{-16}$ level. The detection of air showers
initiated by charged primaries (ultrahigh-energy cosmic-rays) and neutral
primaries (TeV gamma-rays) allows us to obtain new indirect bounds
ranging from the $10^{-15}$ level to the $10^{-19}$ level.
Possible physics implications are briefly discussed.
\end{abstract}

\maketitle

\thispagestyle{fancy}

\section{INTRODUCTION}
\label{sec:introduction}

There is a unique Lorentz-violating (LV) modification of
the Maxwell--Jordan--Pauli theory~\cite{Heitler1954}
of photons, which maintains gauge invariance, \DS{CPT}, and
renormalizability. Restricting this modified-Maxwell theory to the nonbirefringent
sector and adding a standard spin--$\half$ Dirac particle with minimal coupling
to the nonstandard photon, the resulting modified-quantum-electrodynamics model
has 9 dimensionless ``deformation parameters.'' In this contribution, new bounds
are presented, which improve significantly upon current laboratory bounds.

The basic idea~\cite{Beall1970,ColemanGlashow1997} behind these bounds is to
consider novel types of particle decays (absent in the Lorentz-invariant
theory) and to obtain bounds on the LV parameters from the inferred absence
of these decays in air-shower events observed in the Earth's upper
atmosphere.

The present write-up follows the talk given at the conference, but one
crucial result obtained afterwards is added: a lower bound on the isotropic
LV parameter from TeV gamma-rays (Sec.~\ref{subsec:TeV-gamma-ray-bound}).
Natural units with $c=\hbar=1$ are used throughout.

%%\newpage
\section{THEORY}
\label{sec:theory}
\subsection{LV Photon Model}
\label{subsec:LV-photon-model}

For the case of a possible Lorentz noninvariance, the following truism holds
with all force: \emph{it is difficult to discover or bound what is unknown.}
Hence, the need for simple concrete models. Consider, then,  a LV deformation
of quantum electrodynamics (QED):
\beq\label{eq:modQED-action}
S_\text{modQED} = S_\text{modM}+S_\text{standD}\,, \eeq with a
modified-Maxwell term~\cite{ChadhaNielsen1983,ColladayKostelecky1998} and a
standard Dirac term~\cite{Heitler1954}
for a spin--$\half$ particle with charge $e$ and mass $M$:
\begin{subequations}\label{eq:modM-standD-actions}
\beqa \vspace*{-0\baselineskip} S_\text{modM} &=& \int_{\mathbb{R}^4} \id^4 x
\; \Bigg( -\frac{1}{4}\, \Big(\eta^{\mu\rho}\eta^{\nu\sigma}
+\kappa^{\mu\nu\rho\sigma}\Big) \,
\Big( \partial_\mu A_\nu(x) - \partial_\nu A_\mu(x)         \Big)\,
\Big( \partial_\rho A_\sigma(x) - \partial_\sigma A_\rho(x) \Big)
\Bigg)\,,
\label{eq:modM-action} \\
S_\text{standD} &=&\int_{\mathbb{R}^4} \id^4 x \; \overline\psi(x) \Big(
\gamma^\mu \big(\ii\,\partial_\mu -e A_\mu(x) \big) -M\Big) \psi(x)\,.
\label{eq:standD-action} \eeqa
\end{subequations}
The above theory is gauge-invariant, \DS{CPT}--invariant, and power-counting
renormalizable.

In the modified-Maxwell action \eqref{eq:modM-action},
$\kappa^{\mu\nu\rho\sigma}$ is a constant background tensor with $19$
independent components. Ten birefringent parameters are already constrained
at the $10^{-32}$ level~\cite{KosteleckyMewes2002}. Now, set these 10
birefringent parameters to zero, so that 9 nonbirefringent parameters remain
(in the notation of Ref.~\cite{KosteleckyMewes2002}):
\begin{itemize}
\item[3] parity-odd nonisotropic parameters
collected in an antisymmetric traceless $3\times 3$ matrix
$(\widetilde{\kappa}_{\text{o}+})^{mn}\,$; \vspace*{-1mm}
\item[5] parity-even nonisotropic parameters
collected  in a symmetric traceless $3\times 3$ matrix
$(\widetilde{\kappa}_{\text{e}-})^{mn}\,$; \vspace*{-1mm}
\item[1] parity-even isotropic parameter $\widetilde{\kappa}_\text{tr}\,$.
\end{itemize}
The current laboratory bounds
on these 9 parameters will be summarized in Sec.~\ref{sec:current-laboratory-bounds}.

\subsection{LV Particle Decays}
\label{subsec:LV-particle-decays}

The violation of Lorentz invariance in the modified-QED
action \eqref{eq:modQED-action}
leads to modified propagation properties of the photon
(denoted $\widetilde{\gamma}$). The modified photon
propagation, in turn, allows for new types of particle decays.

\begin{figure*}[t]
\centering
\includegraphics[width=80mm]{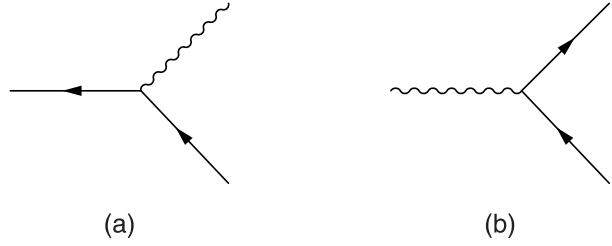}
\vspace*{-2mm} \caption{Feynman diagrams with time running from left to right
and positive charge flowing in the direction of the arrow for (a) the
vacuum-Cherenkov-radiation process and (b) the photon-decay process, both
evaluated for the isotropic Lorentz-violating deformation of quantum
electrodynamics given by
\eqref{eq:modQED-action}--\eqref{eq:modM-standD-actions}.}
\label{fig:Feynman-diagrams}
\end{figure*}

In this contribution, we consider two such decay processes
(Figs.~\ref{fig:Feynman-diagrams}ab), whose occurrence depends on the signs
of the LV parameters, possibly in combination with the flight-direction
vector of the initial particle:
\beq
(a):\;  p^\pm \to p^\pm\,\widetilde{\gamma}
\,,\qquad (b):\;  \widetilde{\gamma} \to p^{-}\,p^{+} \,,
\label{eq:decay-a-b}
\eeq
where $p^\pm$ stands for the electron/positron particle
($e^{-}/e^{+}$ in standard notation) from the original vectorlike $U(1)$
gauge theory, that is, pure QED~\cite{Heitler1954}.
It is also possible to take the charged particles $p^{+}/p^{-}$ in
\eqref{eq:decay-a-b} to correspond to a simplified version of the
proton/antiproton (namely, a Dirac particle with partonic effects neglected
in first approximation). Process (\ref{eq:decay-a-b}a) has been called
``vacuum Cherenkov radiation'' in the literature and process
(\ref{eq:decay-a-b}b) ``photon decay.'' See
Ref.~\cite{KaufholdKlinkhamer2006} for a general discussion of LV decay
processes, starting from Lorentz-noninvariant scalar models.

For the vacuum-Cherenkov process (\ref{eq:decay-a-b}a) in the full
nonbirefringent theory
\eqref{eq:modQED-action}--\eqref{eq:modM-standD-actions}, the square of the
threshold energy is given
by~\cite{Altschul2007,KaufholdKlinkhamer2007,KlinkhamerRisse2008b}
\begin{subequations}\label{eq:Ethreshold-VCR-PD}
\beq \label{eq:Ethreshold-VCR}
\left(E_\text{thresh}^\text{(a)}\right)^2 =
\frac{M^2}{R\Big[\,2\,\widetilde{\kappa}_\text{tr}
-\epsilon^{ijk}\:(\widetilde{\kappa}_{\text{o}+})^{ij}\: \widehat{\vecbf
q}^{k} -(\widetilde{\kappa}_{\text{e}-})^{jk}\:\widehat{\vecbf
q}^{j}\,\widehat{\vecbf q}^{k}\,\Big]} + \text{O}\left(M^2\right), \eeq for
nonbirefringent LV parameters $|\widetilde{\kappa}^{\mu\nu}| \ll 1$ and ramp
function $R[x] \equiv \big(x+|x|\big)/2$.

For the photon-decay process (\ref{eq:decay-a-b}b) in the restricted
isotropic theory with $\widetilde{\kappa}_\text{tr}<0$ and
$(\widetilde{\kappa}_{\text{o}+})^{mn}=(\widetilde{\kappa}_{\text{e}-})^{mn}=0$,
the square of the energy threshold is given by~\cite{KlinkhamerSchreck2008}
\beq\label{eq:Ethreshold-PD} \left(E_\text{thresh}^\text{(b)}\right)^2
=\frac{1-\widetilde{\kappa}_\text{tr}}{-\widetilde{\kappa}_\text{tr}}\;2\,M^2
=\frac{2\, M^2}{-\widetilde{\kappa}_\text{tr}} +2\,M^2\,, \eeq
\end{subequations}
where the last expression holds for all $\widetilde{\kappa}_\text{tr}\in
[-1,0)$ with a well-behaved decay rate
(see Ref.~~\cite{KlinkhamerSchreck2008} for details).

%%\newpage
\section{CURRENT LABORATORY BOUNDS}
\label{sec:current-laboratory-bounds}

The current laboratory bounds on the nonbirefringent parameters of
modified-Maxwell theory are as follows:
\begin{itemize}
\item
direct bounds at the $10^{-12}$ level~\cite{Mueller-etal2007}
for the three
nonisotropic parameters in $\widetilde{\kappa}_{\text{o}+}\,$; \vspace*{-1mm}
\item
direct bounds at the $10^{-14}$ to $10^{-16}$ levels~\cite{Mueller-etal2007}
for the five
nonisotropic parameters in $\widetilde{\kappa}_{\text{e}-}\,$; \vspace*{-1mm}
\item
direct bound at the $10^{-7}$ level~\cite{Reinhardt-etal2007}
for the single
isotropic parameter $\widetilde{\kappa}_\text{tr}\,$; \vspace*{-1mm}
\item
indirect bound at the $10^{-8}$ level~\cite{CaroneSherVanderhaeghen2006} for
$\widetilde{\kappa}_\text{tr}\,$
from the measured value of the electron anomalous moment; %% $a_e \equiv (g_e-2)/2$
\vspace*{-1mm}
\item
indirect bound at the $10^{-11}$ level~\cite{Hohensee-etal2008} for
$\widetilde{\kappa}_\text{tr}\,$ from experiments at particle colliders
(LEP and Tevatron).
\end{itemize}

%%\newpage
\section{NEW INDIRECT BOUNDS}
\label{sec:new-indirect-bounds}

\subsection{Threshold Conditions}
\label{subsec:threshold-conditions}

A remarkable suggestion~\cite{Beall1970,ColemanGlashow1997}
has been made for a way to obtain bounds on nonstandard parameters,
e.g., the 9 Lorentz-violating parameters from Sec.~\ref{subsec:LV-photon-model}
(here, occasionally written as $\widetilde{\kappa}$).
The argument proceeds in three steps:
\vspace*{-0mm}\begin{itemize}
\item
if vacuum Cherenkov radiation or photon decay has a threshold energy
$E_\text{thresh}(\widetilde{\kappa})$, then \mbox{UHECRs} or TeV gamma-rays
with $E>E_\text{thresh}$ cannot travel far, as they rapidly radiate away
their energy or simply disappear;\vspace*{-1mm}\item
this implies that, if an UHECR or TeV gamma-ray of energy $E$ is detected,
its energy must be at or below threshold, $E \leq
E_\text{thresh}(\widetilde{\kappa})$; \vspace*{-1mm}
\item
the last inequality gives, using the thresholds (\ref{eq:Ethreshold-VCR-PD}ab)
for processes (\ref{eq:decay-a-b}ab), an upper bound on the LV parameters,
\beq\label{eq:VCR-PD-conditions} (a):\; R\Big[
\,2\,\widetilde{\kappa}_\text{tr}
-\epsilon^{ijk}\:(\widetilde{\kappa}_{\text{o}+})^{ij}\: \widehat{\vecbf
q}^{k} -(\widetilde{\kappa}_{\text{e}-})^{jk}\:\widehat{\vecbf
q}^{j}\,\widehat{\vecbf q}^{k}\, \Big]\leq M^2/E^2\,, \qquad (b):\;
- \,\widetilde{\kappa}_\text{tr} \leq 2\;M^2/E^2\,,
\eeq with the energy $E$ of the primary,
its flight direction $\widehat{\vecbf q}$, and the
mass $M$ of the Dirac particle involved as input.
\end{itemize}\vspace*{-0mm}
The resulting UHECR/gamma-ray $\widetilde{\kappa}$
bounds depend on the energies and
flight directions of the charged/neutral
pri\-ma\-ries at the top of the Earth's atmosphere.
They are ``terrestrial'' bounds, rather than ``astrophysical'' bounds.

\subsection{Terrestrial UHECR Bounds}
\label{subsec:UHECRs-bounds}

From the absence of the vacuum-Cherenkov process (\ref{eq:decay-a-b}a) for 29
selected UHECR events with primary energies above $57\;\text{EeV}= 5.7\times
10^{19}\;\text{eV}$ (27 events from Auger, 1 event from AGASA, and 1 event
from Fly's Eye), the following two--$\sigma$ ($98 \%\;\text{CL}$) bounds have been
obtained~\cite{KlinkhamerRisse2008b}:
\begin{subequations}\label{eq:SMEbounds-nine}
\begin{eqnarray}
\hspace*{-10mm} (ij)\in \{(23),(31),(12)\} :\;\quad
\big|(\widetilde{\kappa}_{\text{o}+})^{(ij)}\big| &<& 2 \times 10^{-18}\,,
\label{eq:SMEbounds-nonisotropic-odd}\\
\hspace*{-10mm} (kl)\in \{(11),(12),(13),(22),(23)\} :\;\quad
\big|(\widetilde{\kappa}_{\text{e}-})^{(kl)}\big| &<& 4 \times 10^{-18}\,,
\label{eq:SMEbounds-nonisotropic-even}\\
\hspace*{-10mm} \;\hspace*{12.5mm} \widetilde{\kappa}_\text{tr} &<& 1.4
\times 10^{-19}\,, \label{eq:SMEbounds-isotropic-upperbound}
\end{eqnarray}
for a conservative value $M= 56\;\text{GeV}$ in the threshold condition
(\ref{eq:VCR-PD-conditions}a). Note that the quoted confidence level (CL)
percentage holds also for (\ref{eq:SMEbounds-nine}ab), because the bounded
quantities in the analysis were nonnegative (cf. Ref.~\cite{Lyons1986}).

\subsection{Terrestrial TeV Gamma-Ray Bound}
\label{subsec:TeV-gamma-ray-bound}

From the absence of the photon-decay process (\ref{eq:decay-a-b}b)
for $E_{\widetilde{\gamma}}= 30\;\text{TeV}=3.0 \times 10^{13}\;\text{eV}$
gamma-ray photons from a particular supernova remnant observed by the HESS
imaging-atmospheric-Cherenkov-telescope array,
the following two--$\sigma$ ($98 \%\;\text{CL}$) bound has been
obtained~\cite{KlinkhamerSchreck2008}: \beq - 9 \times 10^{-16} <
\widetilde{\kappa}_\text{tr}\,, \label{eq:SMEbounds-isotropic-lowerbound}
\eeq
\end{subequations}
for an electron mass value $M= 0.511\;\text{MeV}$ in the threshold condition
(\ref{eq:VCR-PD-conditions}b). This lower bound obtained from a neutral primary
nicely complements the upper bound
\eqref{eq:SMEbounds-isotropic-upperbound} obtained from a charged primary.

\subsection{Combined Terrestrial and Astrophysical Bound}
\label{subsec:combined-bounds}

From genuinely astrophysical bounds~\cite{KosteleckyMewes2002} on the 10
birefringent modified-Maxwell-theory parameters at the $10^{-32}$ level and
the ``terrestrial'' UHECR/gamma-ray bounds \eqref{eq:SMEbounds-nine} on the 9
nonbirefringent parameters,
the following two--$\sigma$ ($98 \%\;\text{CL}$)
bound is obtained for \emph{all} components of the background tensor
$\kappa^{\mu\nu\rho\sigma}$ in the general action \eqref{eq:modM-action}:
\beq\label{eq:UHECR-kappa-tensor-bound}
\max_{\{\mu,\nu,\rho,\sigma\}} \; |\kappa^{\mu\nu\rho\sigma}|< 5\times 10^{-16} \,,
\eeq
where the fact has been used that the largest entry
of $|\kappa^{\mu\nu\rho\sigma}|$ has a value
$(1/2)\,|\widetilde{\kappa}_\text{tr}|$ if the other 18 parameters are
negligibly small. Remark that the indirect bound
\eqref{eq:UHECR-kappa-tensor-bound} holds in a Sun-centered, nonrotating frame
of reference~\cite{KosteleckyMewes2002}.

%%\newpage
\begin{figure*}[t]\centering
\includegraphics[width=66mm]{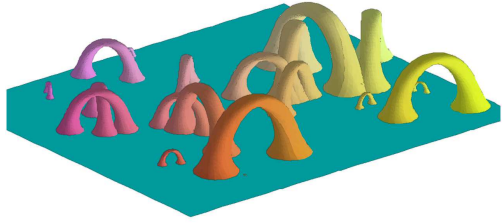}
\vspace*{-4mm} \caption{Sketch of a static classical spacetime-foam
manifold.
\vspace*{-2mm}}
\label{fig:2}
\end{figure*}

\section{DISCUSSION}
\label{sec:discussion}

Explicit calculations~\cite{BernadotteKlinkhamer2007}
of standard photons and standard Dirac
particles propagating over simple classical spacetime-foam manifolds
reproduce, in the large-wavelength limit,
a restricted (isotropic) version of
\mbox{model \eqref{eq:modQED-action}--\eqref{eq:modM-standD-actions}:}
\vspace*{-0mm}
\beq\label{eq:alpha0-calculated}
2\,\widetilde{\kappa}_\text{tr} =
\big(\,\widetilde{b}\,\big/\,\widetilde{l}\;\big)^4 \,,\qquad
(\widetilde{\kappa}_{\text{o}+})^{mn}=(\widetilde{\kappa}_{\text{e}-})^{mn}=0\;,
\eeq
\vspace*{-6mm}\newline
for randomly orientated ``defects'' with an effective size $\widetilde{b}$
and an average separation $\widetilde{l}$ (cf. Fig.~\ref{fig:2}). The
heuristics of the result is well understood, as the type of Maxwell solution
found for the classical spacetime foam is analogous to the solution from the
so-called ``Bethe holes'' for waveguides \cite{Bethe1944}. In both cases, the
standard Maxwell plane wave is modified by the radiation from fictitious
multipoles located in the holes or defects. But there is a difference: for
Bethe, the holes are in a material conductor, whereas for us, the defects are
holes in space itself.

The UHECR bound \eqref{eq:SMEbounds-isotropic-upperbound} then
implies that a single-scale $\big(\widetilde{b}\sim \widetilde{l}\big)$
classical spacetime foam is ruled out. This conclusion holds, in fact, for
arbitrarily small defect size $\widetilde{b}\,$, as long as a classical
\st~makes sense. That is, down to distances at which the classical-quantum
transition occurs, possibly of order
$l_\text{Planck} \equiv  \sqrt{\hbar\,G_\text{N}/c^3}
                 \approx 1.6\times 10^{-35}\,\text{m}$.

This result is really like having a null experiment and there is an analogy
with the Michelson--Morley experiment~\cite{MichelsonMorley1887}:
\emph{theorists expect novel effects which are not seen by experimentalists.}

In turn, this suggests the need for radically new concepts. Then, there was
the  ``relativity of simultaneity'' introduced by
Einstein~\cite{Einstein1905}. Now, for the quantum origin of spacetime,
there is $\ldots$ (alas, the margin is too narrow!) %% Hanc marginis exguitas non caperet [Fermat]
\vspace*{-4mm}

%%\newpage

\end{document}